\documentclass[aps,prl,twocolumn,showpacs,psfig,superscriptaddress,longbibliography]{revtex4-1}
\usepackage{textcomp}
\usepackage{times}
\usepackage{graphicx}
\usepackage{float}
\usepackage{latexsym,amsmath,amssymb,bm,euscript}
\usepackage{color}
\usepackage{subfigure}
\usepackage{epstopdf}
\usepackage[colorlinks=true,linkcolor=blue,citecolor=blue,urlcolor=blue]{hyperref}
\usepackage{hyperref}
\usepackage{soul}
\usepackage[normalem]{ulem}
\usepackage{mathrsfs}
\usepackage{amsmath,amssymb}
\usepackage{lettrine}
\usepackage{xfrac}
\usepackage{xspace}
\usepackage{textcomp}
\usepackage{wasysym}
\usepackage{xr}
\graphicspath{{./Figs/}}

\begin{document}

\title{Thermodynamic Response and Neutral Excitations in Integer and Fractional Quantum Anomalous Hall States Emerging from Correlated Flat Bands}
\author{Hongyu Lu}
\affiliation{Department of Physics and HKU-UCAS Joint Institute
	of Theoretical and Computational Physics, The University of Hong Kong,
	Pokfulam Road, Hong Kong SAR, China}
\author{Bin-Bin Chen}
\affiliation{Department of Physics and HKU-UCAS Joint Institute
	of Theoretical and Computational Physics, The University of Hong Kong,
	Pokfulam Road, Hong Kong SAR, China}
\author{Han-Qing Wu}
\affiliation{Guangdong Provincial Key Laboratory of Magnetoelectric Physics and Devices, School of Physics, Sun Yat-sen University, Guangzhou 510275, China }
\author{Kai Sun}
\email{sunkai@umich.edu}
\affiliation{Department of Physics, University of Michigan, Ann Arbor, Michigan 48109, USA}
\author{Zi Yang Meng}
\email{zymeng@hku.hk}
\affiliation{Department of Physics and HKU-UCAS Joint Institute
	of Theoretical and Computational Physics, The University of Hong Kong,
	Pokfulam Road, Hong Kong SAR, China}
\begin{abstract}
Integer and fractional Chern insulators have been extensively explored in correlated flat band models. Recently, the prediction and experimental observation of fractional quantum anomalous Hall (FQAH) states with spontaneous time-reversal-symmetry breaking have garnered attention. While the thermodynamics of integer quantum anomalous Hall (IQAH) states have been systematically studied, our theoretical knowledge on thermodynamic properties of FQAH states has been severely limited. Here, we delve into the general thermodynamic response and collective excitations of both IQAH and FQAH states within the paradigmatic flat Chern-band model with remote band considered. Our key findings include: 
i) In both $\nu=1$ IQAH and $\nu=1/3$ FQAH states, even without spin fluctuations, the charge-neutral collective excitations would lower the onset temperature of these topological states, to a value significantly smaller than the charge gap, due to band-mixing and multi-particle scattering;
ii) By employing large-scale thermodynamic simulations in FQAH states in the presence of strong inter-band mixing between $C=\pm1$ bands, we find that the lowest collective excitations manifest as the zero-momentum {\it excitons} in the IQAH state, whereas in the FQAH state, they take the form of {\it magneto-rotons} with finite momentum; iii) The unique charge oscillations in FQAH states are exhibited with distinct experimental signatures, which we propose to detect in future experiments.  
\end{abstract}

\date{\today }
\maketitle

\noindent{\textcolor{blue}{\it Introduction.}---} 
 The intricate interplay between electronic interaction and band topology can give rise to exotic  quantum states of matter. A well-known example is the quantum anomalous Hall (QAH) effect, emerging at zero magnetic field due to intrinsic ferromagnetism~\cite{Haldane1988_qah, CZChang2013, CXliu2016_qah, XLQi2020_qhe}. Beyond the integer case, fractional QAH (FQAH) states can be even more intriguing, which have been theoretically predicted in a class of topological flat band models without Landau levels, known as fractional Chern insulators (FCI) \cite{tangHigh2011, KSun2011_model, Neupert2011_fci, DNSheng2011_fci, regnaultFractional2011}. The ground-state phase diagrams of these models have been extensively studied~\cite{DNSheng2011_fci, regnaultFractional2011, DNSheng2011_bosonfci, WZhu2015_nonabelian, WZhu2016_nonabelian, TSZeng2017_fqah, TSZeng2018_SUn}, and the connections between FQH and FCI (FQAH) states are discussed~\cite{YHWu2012fci, YLWu2012fci, Scaffifi2012_adiabatic, Repellin2014_roton, andrews2023stability}. Besides, the emergence of semiconductor moir\'e materials~\cite{makSemiconductor2022}, supporting tunable correlated topological flat bands~\cite{MacDonald2011, liSpontaneous2021, ashvin2021, BBChen2021_TBG, XYLin2022_exciton, GPPan2023_qah, zhangPolynomial2023, huangEvolution2023, Cano2023_fci}, is expected to offer ideal opportunities for realizing FQAH states~\cite{wuTopological2019, yuGiant2019, wang2023fractional, liSpontaneous2021, devakul2021magic,Fu2021_fci, dong2023fqah_graphene, zhou2023fqah_graphene}.

Recent advancements provide substantial support for this  perspective. Firstly, the experimental demonstration of QAH states in various systems at zero magnetic field has been achieved ~\cite{CZChang2013, CXliu2016_qah, MSerlin2020_qah, Chen2020_qah, YBZhang2020_qah, KFMak2021_qah}. Additionally, FCIs have been observed in graphene-based systems under a finite magnetic field~\cite{ericObservation2018, xieFractional2021}. Most recently, experiments have reported evidence of FQAH states with zero external magnetic field, in  twisted molybdenum ditelluride (MoTe$_2$) bilayers~\cite{caiSignature2023, zengThermodynamic2023, xu2023_fci, park2023_fqah} and in rhombohedral pentalayer graphene/hBN moir\'e superlattices~\cite{ZLu2023fqh_graphene}. 

An intriguing aspect of these FCI experiments is the relatively high onset temperature of Hall plateaus, compared to some FQHl experiments. A key question for future investigations revolves  understanding the physics that dictates this onset temperature and exploring the possibility of pushing it to even higher values, potentially reaching room-temperature scales~\cite{tangHigh2011}. Notably, in MoTe$_2$~\cite{caiSignature2023, park2023_fqah}, there exists intense spin fluctuations above the onset temperature, which could potentially affect the Hall transport at temperature much lower than the charge gap. However, the charge fluctuations have not been well considered and they should be the key factor of whether the onset temperature could be largely enhanced up to the charge-gap level if there exist no spin fluctuations.

For the IQAH state in twisted bilayer graphene at odd integer fillings~\cite{MSerlin2020_qah,xieFractional2021,stepanovCompeting2021}, the charge-neutral exciton was found to acquire a small gap that determined the transition temperature of the IQAH, much lower (one magnitude) than the charge gap of the insulator ground state~\cite{XYLin2022_exciton,GPPan2023_qah,zhangPolynomial2023,huangEvolution2023}. However, for FQAH states, it's crucial to address a significant knowledge gap between theoretical and experimental studies. While experiments are inherently conducted at finite temperatures, the majority of numerical and theoretical investigations of FQAH states focus on zero temperature.

Although charge-neutral collective excitations such as magneto-roton and chiral graviton have been intensively discussed in FQH systems~\cite{Haldane1985_roton,girvinMagneto1986,kangInelastic2000,kukushkinDispersion2009,Haldane2019_graviton,Haldane2021_graviton,YZWang2022_graviton, Haldane2022_roton,KYang2022_roton, YZWang2023_graviton}, these are basically at zero temperature, our understanding of thermodynamic properties, especially those near or above the onset temperature is exceptionally limited. But since the understanding of finite temperature properties in both IQAH and FQAH is vital for unraveling the factors to promote the FCI physics to higher temperatures, there is an urgent need to address such fundamental problem.

\begin{figure}[htp!]
	\centering	
		\includegraphics[width=0.5\textwidth]{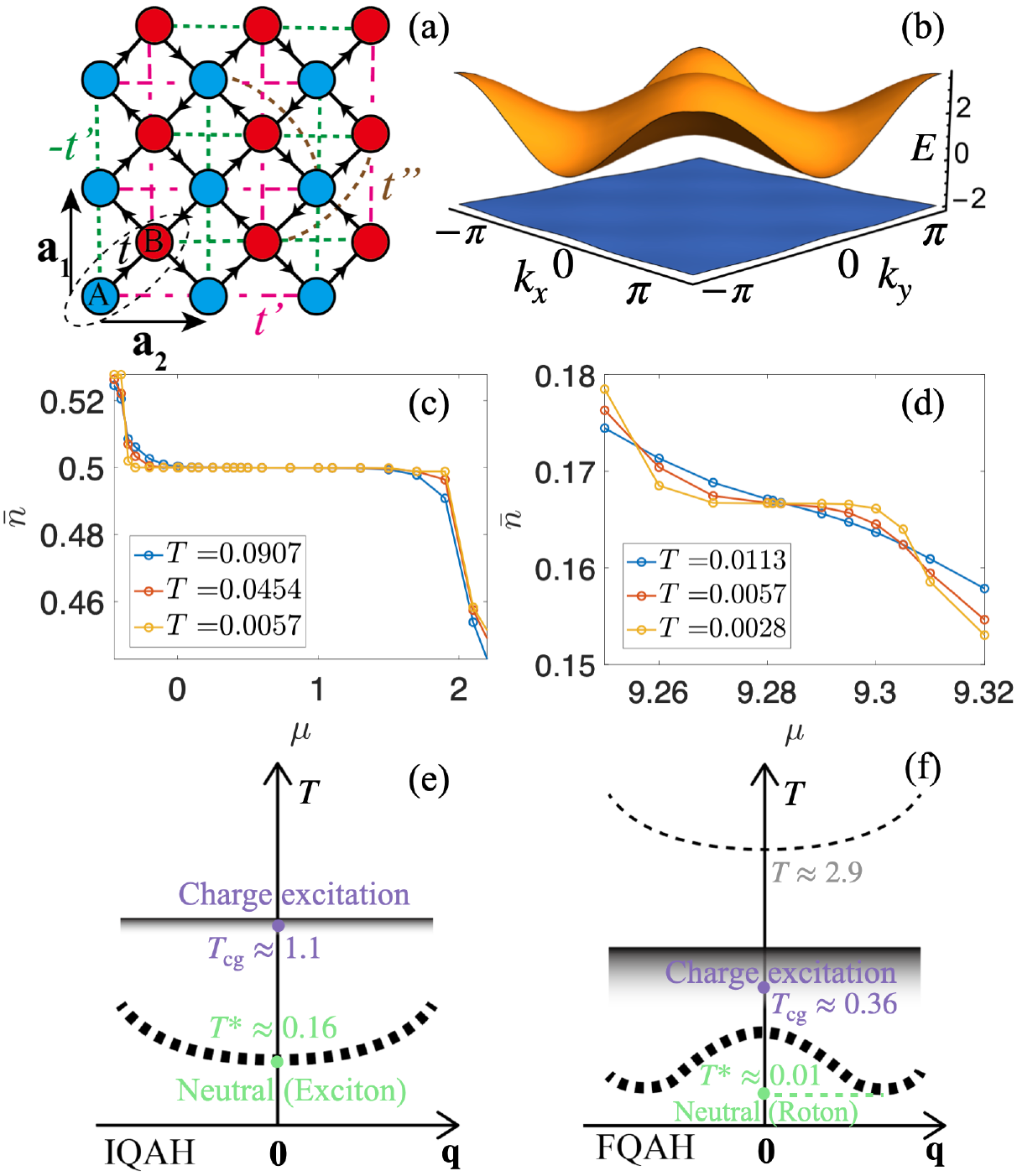}
	\caption{\textbf{Lattice model, correlated flat Chern bands, $\bar n-\mu$ plateaus, and the separation of energy scales}. (a) Checkerboard lattice with different hoppings denoted by colors; and lattice primitive vectors are $\mathbf{a_1}= (0,1)$, $\mathbf{a_2}=(1,0)$. The arrows represent the directions of the loop current. (b) Band dispersion of the tight-binding Hamiltonian, with the lower band being nearly flat. Average density ${\bar n}$ versus $\mu$ at different temperatures for the $\nu=1$ plateau at $V_1=1$ (c), and $\nu=1/3$ plateau at $V_1=4$ (d). (e) Schematic excitation spectra for $\nu=1$ IQAH. The lowest charge-neutral excitations with zero momentum $\sim T^{\ast}$, is much lower than the charge gap scale $\sim T_\mathrm{cg}$. (f) Schematic excitation spectra for $\nu=1/3$ FQAH. A finite-momentum roton mode $\sim T^{\ast}$ is found to be the dominant, at temperature scales below the charge gap. At $T$ much higher than the charge gap, another zero-momentum charge-neutral excitation is observed.}
	\label{fig_fig1}
\end{figure}

In this letter, we study the general properties of both IQAH and FQAH states from the thermodynamic perspective in the paradigmatic correlated flat band model with spinless fermions~\cite{KSun2011_model,DNSheng2011_fci,Neupert2011_fci} in Fig.~\ref{fig_fig1} (a-b). 
Employing the state-of-the-art exponential tensor renormalization group (XTRG)~\cite{BBChen2018_XTRG}, complemented with density matrix renormalization group (DMRG) and  exact diagonalization (ED).
We find that, i) in both $\nu=1$ IQAH and $\nu=\frac{1}{3}$ FQAH states, even without spin fluctuations, the charge-neutral collective excitations in spinless fermions would lower the onset temperature (Hall plateau) of these topological states, much lower than the charge-gap scale; ii) the low-energy charge-neutral collective excitations are excitons in the IQAH state, manifested as density fluctuations with zero momentum transfer; in contrast, the corresponding excitations in FQAH states are magneto-rotons, signified as finite-momentum peaks  in the density fluctuations. 
iii) These roton excitations lead to unique real-space patterns in density-density correlations, which could be observed in experiments using local probes in analogy to Friedel oscillations.

\noindent{\textcolor{blue}{\it Model and Method.}---} We study the on the checkerboard lattice (Fig.~\ref{fig_fig1} (a)), 
\begin{equation}
	\begin{aligned}
	H =&-t\sum_{\langle i,j\rangle}e^{i\phi_{ij}}(c_i^\dagger c^{\ }_j+h.c.)-\sum_{\langle\langle i,j \rangle\rangle}t'_{ij}(c_i^\dagger c^{\ }_j+h.c.)\\
	&-t''\sum_{\langle\langle\langle i,j \rangle\rangle\rangle}(c_i^\dagger c^{\ }_j+h.c.)+V_1\sum_{\langle i,j\rangle}(n_i-\frac{1}{2})(n_j-\frac{1}{2})
	\end{aligned}
\end{equation} 
with nearest-neighbor (NN, $t$), next-nearest-neighbor (NNN, $t'$), and next-next-nearest-neighbor (NNNN, $t''$) hoppings, and NN repulsive interaction ($V_1$). We allow the NN hoppings to carry nonzero complex phase. We are using dimensionless parameters by setting $t=1$
and other tight-binding parameters are: $t'_{ij}=\pm 1/(2+\sqrt{2})$, $t''=-1/(2+2\sqrt{2})$ and $\phi_{ij}=\pm\frac{\pi}{4}$.
 At the noninteracting limit, the dispersions carry Chern number, $C=\pm 1$ for the flat and the remote bands and give rise to the QAH state~\cite{Haldane1988_qah}. We work in the parameter regime that the flat-band width $W$, the gap between the flat and remote band $\Delta$ and the interacting strength $V_1$ 
are chosen such that $W(=0.08)\ll\Delta(=2.34)\sim V_1$. Therefore, the band-mixing effect is well included.

We employ XTRG for finite-temperature simulations and implement the charge $U(1)$ symmetry based on the QSpace framework~\cite{AW2012_QSpace} with up to $D=800$ bond states kept, ensuring the maximum truncation error below $10^{-4}$. For complementary, we also utilized ED and DMRG to obtain ground state properties~\cite{suppl}.

For XTRG, we operate in the grand canonical ensemble by introducing a chemical potential term $H_\mu=\mu\sum_i (\hat n_i-\frac{1}{2})$ to adjust the particle numbers $N_e=\sum_i \langle \hat n_i\rangle_\beta$ (here, $\langle\cdot\rangle_\beta$ denotes the ensemble average at inverse temperature $\beta\equiv\tfrac{1}{T}$) and the system is half-filled without $H_\mu$. 
We denote the total number of lattice sites as $N=N_y\times N_x \times 2$, with $N_y$/$N_x$ representing the number of unit cells along the ${\bf a_1}$/${\bf a_2}$ direction, and we consider a $3\times12\times2$ cylinder in thermodynamic simulations. 
The filling of the flat band $\nu$ and the average density $\bar n$ are given by $\bar n=N_e/N = \nu/2$.

\begin{figure}[tp!]
	\centering		
	\includegraphics[width=0.5\textwidth]{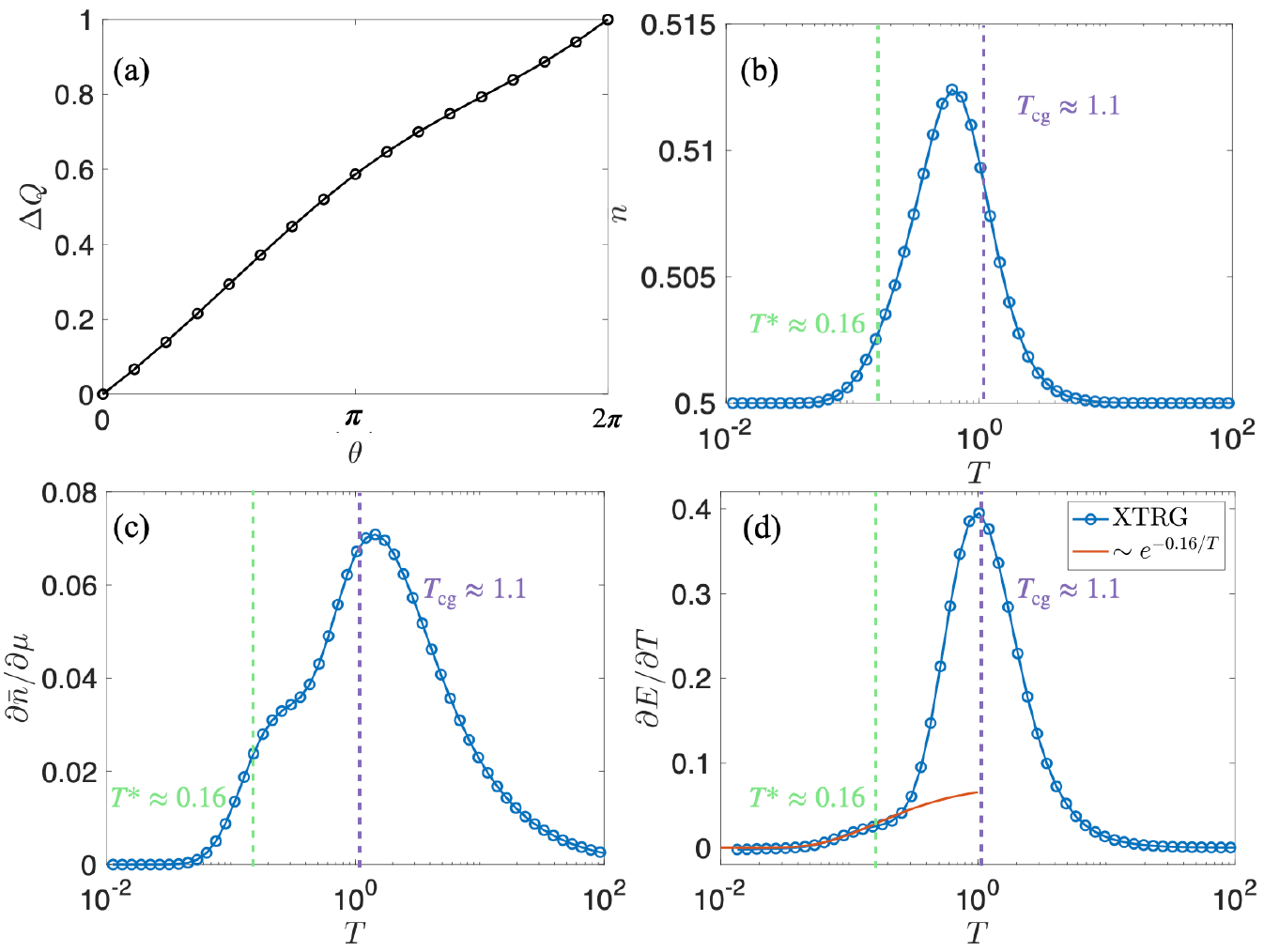}
	\caption{ \textbf{Hall conductivity and thermodynamics of the IQAH state.} (a) Charge pumping. (b-d) Temperature dependence of (b) average density, (c) compressibility, and (d) specific heat. The red solid line in (d) depicts the low-$T$ behavior fitted to $e^{-0.16/T}$. The green and purple dashed lines represent the onset temperature of the QAH effect $T^\ast\approx0.16$ and the charge gap scale $T_{\mathrm{cg}}\approx1.1$.}
	\label{fig_integer}
\end{figure}

\noindent{\textcolor{blue}{\it $\nu=1$ IQAH.}---} Here we focus on flat Chern bands with $C=1$ at filling $\nu=1$. We set $V_1=1$ and first examine the ground state to validate its nontrivial topology, since large $V_1$ triggers a quantum phase transition toward a topologically trivial nematic insulator~\cite{KSun2009_QBT, HQWu2016_QAH, TSZeng2018_QBT, Shouvik2018_QAH, HYLu2022_QAH, HYLu2023QBT}.
The ED simulations of a $4\times 4 \times 2$ torus reveal a non-degenerate  ground state with $C=1$~\cite{suppl} and we utilize DMRG to compute the Hall conductance in a $N_y=3$ cylinder. As depicted in Fig.~\ref{fig_integer} (a), upon inserting one flux quanta, $\theta$ from $0$ to $2\pi$, a quantized charge $\Delta Q=1$ is pumped across the cylinder, indicating a quantized Hall conductance of $\sigma_{xy} = \tfrac{e^2}{h}$~\cite{laughlin1981}.

We then implement thermodynamic simulations. In Fig.~\ref{fig_fig1}~(c), at low temperature, as the chemical potential $\mu$ varies, we observe a plateau at $\bar n=\nu/2=0.5$, which hallmarks a non-compressible state at low temperature, and the estimated charge gap is $\Delta_\mathrm{cg}\approx 1.1$ from the change of $H_\mu$ on the plateau. Subsequently, 
 we fix $\mu$ and explore the temperature dependence of $\bar n$, compressibility $\frac{\partial \bar n }{\partial \mu}$, and specific heat $\frac{\partial E}{\partial T}$ [Fig.~\ref{fig_integer} (b), (c), and (d)]. The specific heat reveals two discernible temperature scales: the charge-gap scale $T_\text{cg}\approx 1.1$, indicated by the prominent peak, and a low-energy scale denoted by the specific heat shoulder near $T^\ast\approx 0.16$. Importantly, below $T^\ast\approx 0.16$, the average density swiftly converges to $\bar n=0.5$, while the compressibility undergoes a rapid decline toward 0 and the specific heat begins to exhibit activation behavior $\sim e^{-T^\ast/T}$. These behaviors indicate the incompressible nature of the QAH state, and thus $T^\ast$ represents the crossover temperature marking the onset of QAH state.
The existence of two distinct temperature scales here is reminiscent of the QAH state in TBG-related models~\cite{XYLin2022_exciton, GPPan2023_qah, zhangPolynomial2023, huangEvolution2023}.

\begin{figure}[htp!]
	\centering		
	\includegraphics[width=0.5\textwidth]{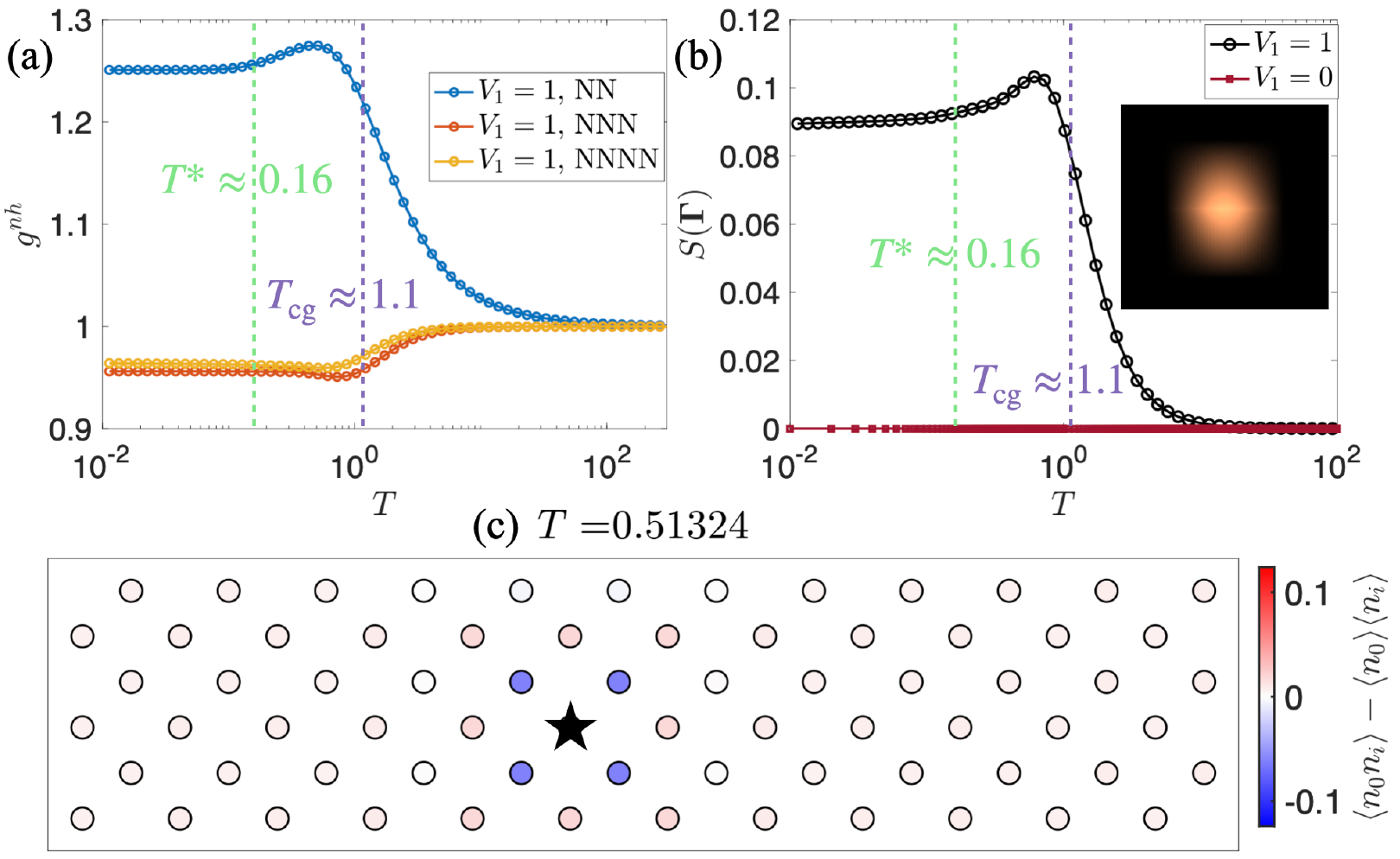}
	\caption{\textbf{Charge-neutral excitations in IQAH.} (a) Pairs of particle-hole correlations with the reference site chosen in the bulk. (b) The temperature dependence of the structure factor at $\Gamma$, in comparison to the non-interacting limit. The inset shows the momentum dependence of the structure factor at $T=0.51324$, which peaks at $\Gamma$. (c) Density-density correlations $\langle n_0n_j \rangle-\langle n_0 \rangle\langle n_j \rangle $, where the site $0$ is denoted by the black asterisk. The correlation is spatially confined.
	}
	\label{fig_integer_excitation}
\end{figure}

The observation of $T^\ast \ll T_\mathrm{cg}$ suggests the existence of low-energy charge-neutral excitations, with a gap much smaller than the charge gap $\Delta_\mathrm{cg}$. To elucidate the nature of these charge-neutral excitations, we examine particle-hole excitations by computing density fluctuations and their Fourier transform $S(\mathbf{q})=\sum_j e^{-i\mathbf{q}(\mathbf{r_0}-\mathbf{r_j})}( \langle n_0n_j\rangle-\langle n_0 \rangle\langle n_j \rangle)$.
As illustrated in the inset of Fig.~\ref{fig_integer_excitation} (b), the fluctuations peak at the $\Gamma$ point, indicating that these charge-neutral excitations predominantly have zero (small) momentum. From the temperature dependence shown in Fig.~\ref{fig_integer_excitation} (b), these excitations at  $\Gamma$ indeed begin to develop at $T< T_\mathrm{cg}$, below the charge-gap scale.
For comparison, we also plot $S(\Gamma)$ in the non-interacting limit ($V_1=0$), where such charge-neutral excitations are absent at $T<T_\mathrm{cg}$, indicating that excitations observed here are a distinctive and interaction-driven feature of correlated flat band systems.

Additionally, we compute the particle-hole correlations $g^{nh}=\frac{\langle \hat n_0 \hat h_i\rangle_\beta}{\langle \hat n_0\rangle_\beta\langle \hat h_i\rangle_\beta}$ between a fixed site $0$ and other sites $i$, where $\hat{h}_i\equiv1-\hat{n}_i$ represents the number of holes. In Fig.~\ref{fig_integer_excitation}(a), above $T^\ast$, the NN particle-hole bunching effect rapidly establishes (while the NNN and NNNN particle-hole correlations exhibit repelling behavior), reaching a maximum around $T\sim 0.51$, suggesting the proliferation of zero-momentum excitons in this temperature regime, and eventually smearing out above the charge excitation scale $T_\text{cg}$.

\noindent{\textcolor{blue}{\it $\nu=1/3$ FQAH.}---}  
We now extend our investigation to the thermodynamics of the $\nu=\frac{1}{3}$ FQAH state, using $V_1=4$ as an example. 
The ground state of this model has been shown to be a FQAH state~\cite{KSun2011_model, DNSheng2011_fci}.
As a benchmark, we show ED results in the SM~\cite{suppl}, which confirm that the Chern number of each of the three degenerate ground states is 1/3.

\begin{figure}[tp!]
\centering		
\includegraphics[width=0.5\textwidth]{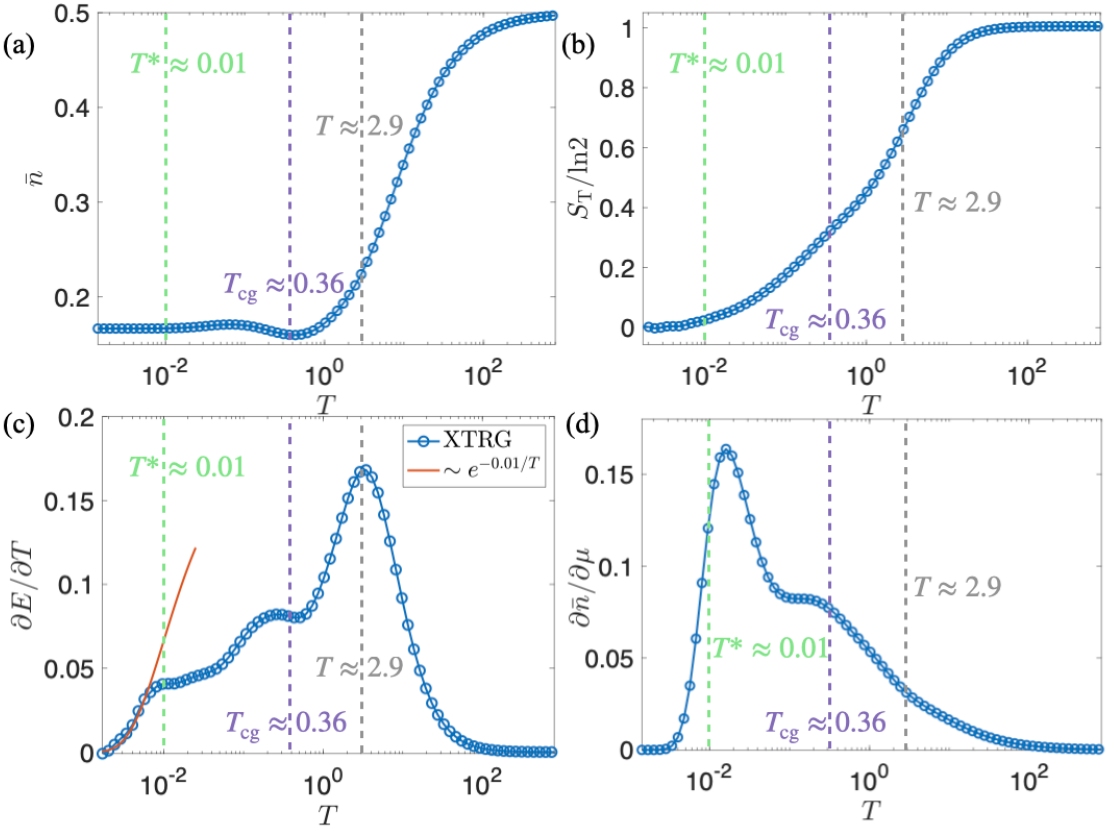}
\caption{\textbf{Thermodynamics of the $\nu=1/3$ FQAH states with $V_1=4$}. 
(a) Average density, (b) thermal entropy, (c) specific heat with the low-$T$ behavior fitted with $e^{-0.01/T}$ (the red solid line) and (d) compressibility, versuse $T$. The three dashed lines represent the roton scale $T^*\approx 0.01$ (green), the charge excitation gap $T_{cg}\approx 0.36$ (purple) and the high-$T$ particle-hole excitation scale $T\approx 2.9$ (grey).}
	\label{fig_frac_ed_thermal}
\end{figure}

For thermodynamic properties, we observe a $\bar n-\mu$ plateau shown in Fig.~\ref{fig_fig1} (d) as well, with the charge gap $\Delta_\text{cg}\simeq 0.36$ approximated from the low-temperature plateau. Setting $\mu=9.281$, we compute the temperature-dependence of $\bar n$, thermal entropy $S$, specific heat $\frac{\partial E}{\partial T}$, and compressibility $\frac{\partial {\bar n}}{\partial T}$ as shown in Fig.~\ref{fig_frac_ed_thermal} (a-d).

From the specific heat, we can identify three temperature scales. The first (lowest) one is a crossover temperature $T^\ast\approx0.01$, situated near the lowest shoulder of the specific heat. Below $T^\ast$, the average density $\bar n$ converges to $1/6$ ($\nu =1/3$), and both thermal entropy $S_T$ and compressibility rapidly approach 0. In addition, for $T<T^\ast$, we observe activation behavior  in specific heat $\frac{\partial E}{\partial T}\sim e^{-T^\ast/T}$. 
Same as the $\nu=1$ case discussed above, this $T^\ast$ marks the on-set of incompressibility of FQAH effect. As will be discussed below, this $T^\ast$ comes from the energy scale of roton excitations.
Above $T^\ast$, the specific heat shows two more features: (1) a hump at $T_\text{cg}\approx0.36$, which is the charge-gap energy scale, and a peak at $T\approx2.9$, which will be linked to a high-energy zero-momentum charge-neutral excitations.

\begin{figure}[htp!]
	\centering		
	\includegraphics[width=0.5\textwidth]{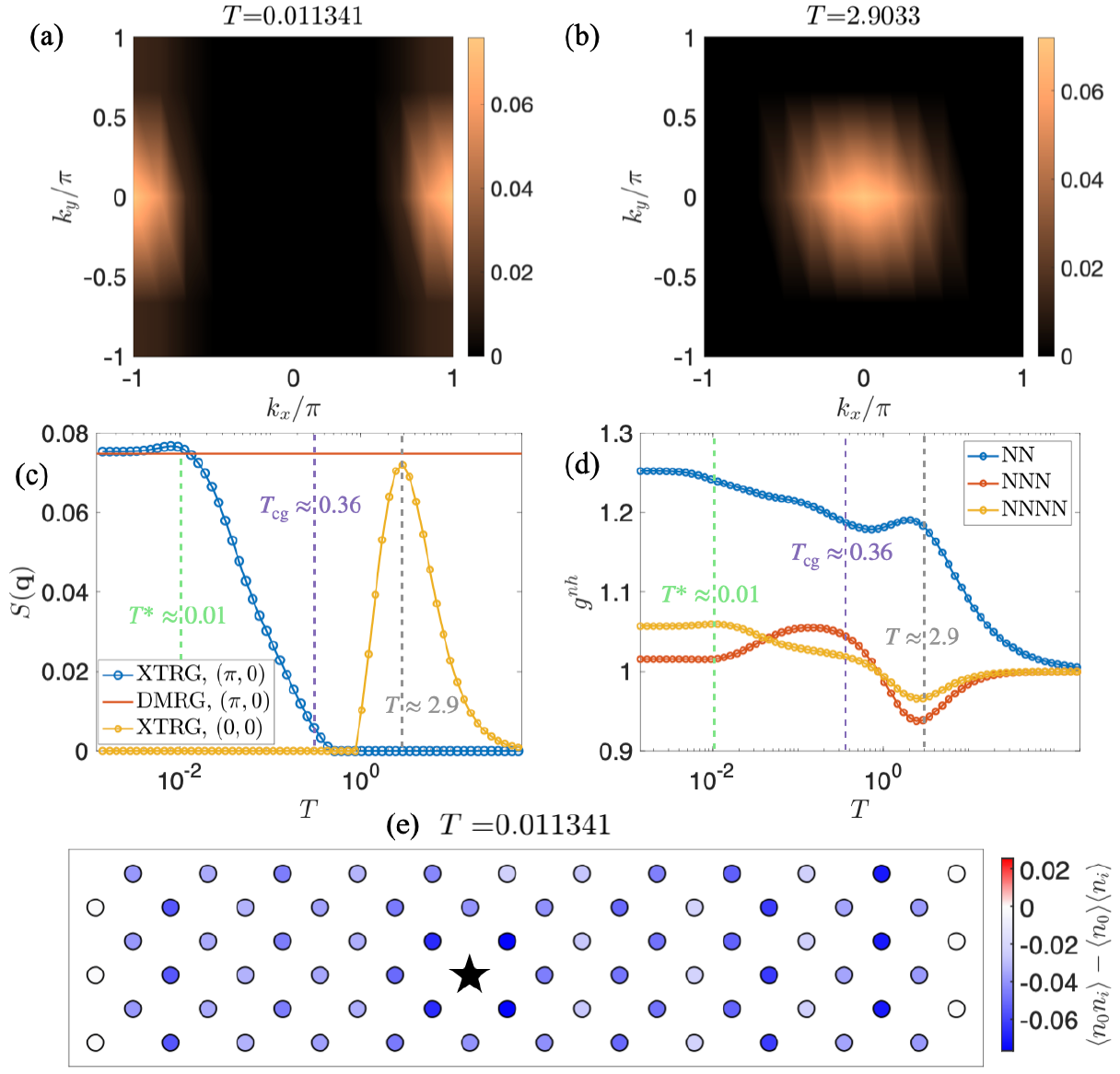}
	\caption{\textbf{Roton excitations in $\nu=1/3$ FQAH}. (a) Structure factor $S(\mathbf{q})$ around $T^\ast$. (b) Structure factor $S(\mathbf{q})$ at $T\approx2.9$.  (c) Temperature dependence of  $S(\mathbf{q})$. The horizontal line marks the $T=0$ value of $S$ from DMRG. (d) Temperature dependence of particle-hole correlations for NN, NNN, and NNNN pairs. (e) Real-space density-density correlation function near $T^\ast$, and the black asterisk represents the reference site. The rotons are spatially extended and exhibit real space oscillations. }
	\label{fig_frac_excitation}
\end{figure}

Likewise, we also observe $T^\ast\ll T_\mathrm{cg}$ here, signifying the presence of low-energy charge-neutral excitations below the charge gap, which lower the onset temperature of Hall transport. Notably, in contrast to the integer case, these charge-neutral excitations in the FQAH phase carry finite momentum, and thus are identified as rotons.
 In Fig.~\ref{fig_frac_excitation} (a), we display the density structure factor $S(\mathbf{q})$ near $T^\ast$, where the highest values are prominently located at $\mathbf{q}=(\pi,0)$. The absence of four-fold rotational symmetry here is due to the geometry, as details shown in the SM~\cite{suppl}. The structure factors at higher $T=2.9033$ are also presented in Fig.~\ref{fig_frac_excitation} (b), with the peak observed at $\Gamma$.
Furthermore, we illustrate $S(\mathbf{q})$ versus $T$ for $\mathbf{q}=(\pi,0)$ and $\Gamma$ in Fig.~\ref{fig_frac_excitation} (c). We observe that the value of $S(\pi,0)$ peaks around $T^\ast$ and saturates to the DMRG result at $T=0$. In contrast, $S(\Gamma)$ remains featureless until reaching much higher temperature $T\approx2.9$, and thus might include particle-hole excitations between the $C=\pm1$ bands. These observations clearly indicate that collective excitation in the FQAH phase  fundamentally  differs from that in the IQAH case. When the value of $S(\mathbf{q})$ is too small, it is hard to numerically obtain accurate and valid values, so we take those extremely small values affected by numerical noise as $0$.

Additionally, pairs of particle-hole correlations $g^{nh}$ are depicted in Fig.~\ref{fig_frac_excitation} (d). At $T\sim 2.9$, similar to the IQAH case, NN sites exhibit particle-hole correlations ($g^{nh}>1$), while NNN and NNNN sites display particle-particle correlations ($g^{nh}<1$). However, the zero-momentum exciton mode here appears only at high temperatures above $T_\text{cg}\approx0.36$. 

As shown in Fig.~\ref{fig_frac_excitation}(e), near $T^\ast$, density-density correlation exhibits a clear periodic oscillation in real space. This pattern is the direct manifestation of the finite-momentum charge-neutral excitations discussed above. 
In analogy to the Friedel oscillations in metals, near a quenched charge impurity in a FQAH state, these roton excitations will generate real-space oscillations in electron density, with the same pattern shown in Fig.~\ref{fig_frac_excitation}(e). Such real-space density oscillations can be directly imaged using STM and/or other local probes.  In contrast, for IQAH states, although low-energy charge-neutral excitations proliferate at $T\sim T^\ast$, due to their small (zero) momentum, real-space oscillations (beyond the lattice periodicity) will not arise around charge impurities, as shown in Fig.~\ref{fig_integer_excitation} (c). The charge oscillation near impurity is similar to early studies in fractional quantum Hall system (under magnetic field)~\cite{FCZhang1985_impurity, Rezayi1985_impurity}, and we note very recently, this feature in FQAH system has also been theoretically proposed in accordance with our work~\cite{song2023halo}.

We emphasize here why the charge-neutral excitations can affect the onset temperature of the Hall plateaus. There are two key factors and the first key physical process here is the multiple-particle scattering. 
Due to energy mismatch, if a charged particle in the ground state absorb multiple charge-neutral excitations, the total energy from these neutral excitations would be enough to create charge excitations and thus change the transport coefficient. The probability of absorbing $n$ excitations is proportional to $\rho^n$, where $\rho$ is the density of charge-neutral excitations. As we increase temperature $T$, $\rho$ increases exponentially. And this probability also increases exponentially, making it easier to create charged excitations. 
This is exactly why we see shrunken density plateau and finite compressibility at $T\sim T^\ast\ll T_{\mathrm{cg}}$ as shown in Fig.\ref{fig_fig1}(c-d), Fig.\ref{fig_integer}(c) and Fig.\ref{fig_frac_ed_thermal}(d), which we believe is the main reason of onset temperature $T^\ast\ll T_{\mathrm{cg}}$ in the FQAH case.
The second factor could be the band-mixing, which is in contrast to the Laudau level physics. In the IQAH state, the $C=1$ band is fully occupied at ground state, and thus the particle-hole excitations between the two $C=\pm1$ Chern bands naturally affect the Hall conductance~\cite{XYLin2022_exciton,GPPan2023_qah,zhangPolynomial2023,huangEvolution2023}

\noindent{\textcolor{blue}{\it Discussions.}---}
In conlusion, we numerically study the generic thermodynamic properties of $\nu=1$ IQAH and $\nu=1/3$ FQAH states, and we note such study is rare, even in the FQH literature. We find the zero-momentum exciton in IQAH state and the finite-momentum roton in FQAH state, as the low-energy excitations that could determine their onset temperature $T^*$, much lower than their zero-temperature charge gap scale $T_{cg}$. This could provide broader perspective to recent experiments where the spin fluctuations are thought to lower the onset temperature much below the charge gap, while our results show that, even without magnon excitations, the onset temperature is still much lower than charge gap in the presence of lower-energy neutral excitations.

Although the neutral gap and charge gap are close in conventional FQH states since the composite fermions are very weakly coupled (while strongly coupled in FCI)~\cite{Wu2012adiabatic}, there also exist FQH states with quite small neutral gap~\cite{Han2023FQH} or even gapless neutral excitation in nematic FQH states~\cite{xia2011nematic,you2014nematic, regnault2017_nematic_fqh, BoYang2020nematic, pu2024nematicFQH}.
Since the study of thermodynamic response in such systems is limited, it would be interesting to numerically study the interplay of thermal fluctuations and neutral excitations.

We also provide experimental signatures to directly probe these finite momentum roton excitations, as an unique feature in FQAH systems.
Moreover, we notice quantum geometry has been studied to be closely related to the critical temperature of superconductors~\cite{Peotta2022_geometry, xuSuperconductivity2022, Bernevig2022metric, peotta2023geometry, Torma2023quantumgeometry}, and it would be interesting to study the interplay of thermal fluctuations and quantum geometry in FCIs.


\begin{acknowledgments}
{\it Acknowledgments}\,---\,
HYL, BBC and ZYM would like to thank Bo Yang, Wang Yao, Wei Zhu and Tian-Sheng Zeng for helpful discussions. HYL, BBC and ZYM acknowledge the support from the Research Grants Council (RGC) of Hong Kong Special Administrative Region of China (Project Nos. 17301721, AoE/P-701/20, 17309822, HKU C7037-22GF, 17302223), the ANR/RGC Joint Research Scheme sponsored by RGC of Hong Kong and French National Research Agency (Project No. A\_HKU703/22). We thank HPC2021 system under the Information Technology Services and the Blackbody HPC system at the Department of Physics, University of Hong Kong, as well as the Beijng PARATERA Tech CO.,Ltd. (URL: https://cloud.paratera.com) for providing HPC resources that have contributed to the research results reported within this paper.
H.Q. Wu acknowledge the support from Guangzhou Basic and Applied Basic Research Foundation (No. 202201011569).
The ED calculations reported were performed on resources provided by the Guangdong Provincial Key Laboratory of Magnetoelectric Physics and Devices, No. 2022B1212010008.
\end{acknowledgments}

\bibliographystyle{apsrev4-2}



\newpage\clearpage
\renewcommand{\theequation}{S\arabic{equation}} \renewcommand{\thefigure}{S%
	\arabic{figure}} \setcounter{equation}{0} \setcounter{figure}{0}

\begin{widetext}
	
	\section{Supplemental Materials for \\[0.5em]
		THERMODYNAMIC RESPONSE AND NEUTRAL EXCITATIONS IN INTEGER AND FRACTIONAL QUANTUM ANOMALOUS HALL STATES EMERGING FROM CORRELATED FLAT BANDS
	}
	In this Supplemental Materials, we introduce the ED calculations with more spectra data, Brillioun zone for different clusters with torus geometry, and the structure factor of density correlations for $\nu=1/3$ FQAH state in section I. In section II, we show the structure factor of density correlations for $\nu=1/3$ FQAH state from DMRG simulations. In section III, we briefly summarize the XTRG method.
	
	\subsection{Section I: ED results of $\nu=1/3$ FQAH state}
	
	The exact diaognalization method is performed using charge $U(1)$ and translational symmetries. Using the charge $U(1)$ symmetry, we can keep the number of electrons fit at certain filling. Using the translational symmetry, we can further reduce the dimension of Hamiltonian matrix. For example, for $\nu=1/3$ and $6\times 4\times 2$ system size, the largest block dimension of Hamiltonian matrix is 15724214. It can be easy to get the ground state using Lanczos method with sparse matrix storage and parallel computing. For calculating the low-energy excited states, we can let each new Lanczos vector is explicitly orthogonalized with respect to all previous ground-state and excited state wavefunctions.
	
	The spectra of different size at $V_1=4$ are shown in Fig.\ref{figsm_ed_twist}(a), with spectra flow shown in Fig.\ref{figsm_ed_twist}(b).
	The 3-fold degenerate ground state are at different momenta for $N=24,30,48$ while they are at the same point for $N=36$, which supports the FQAH ground state.

	\begin{figure}[htp!]
		\centering		
		\includegraphics[width=0.5\textwidth]{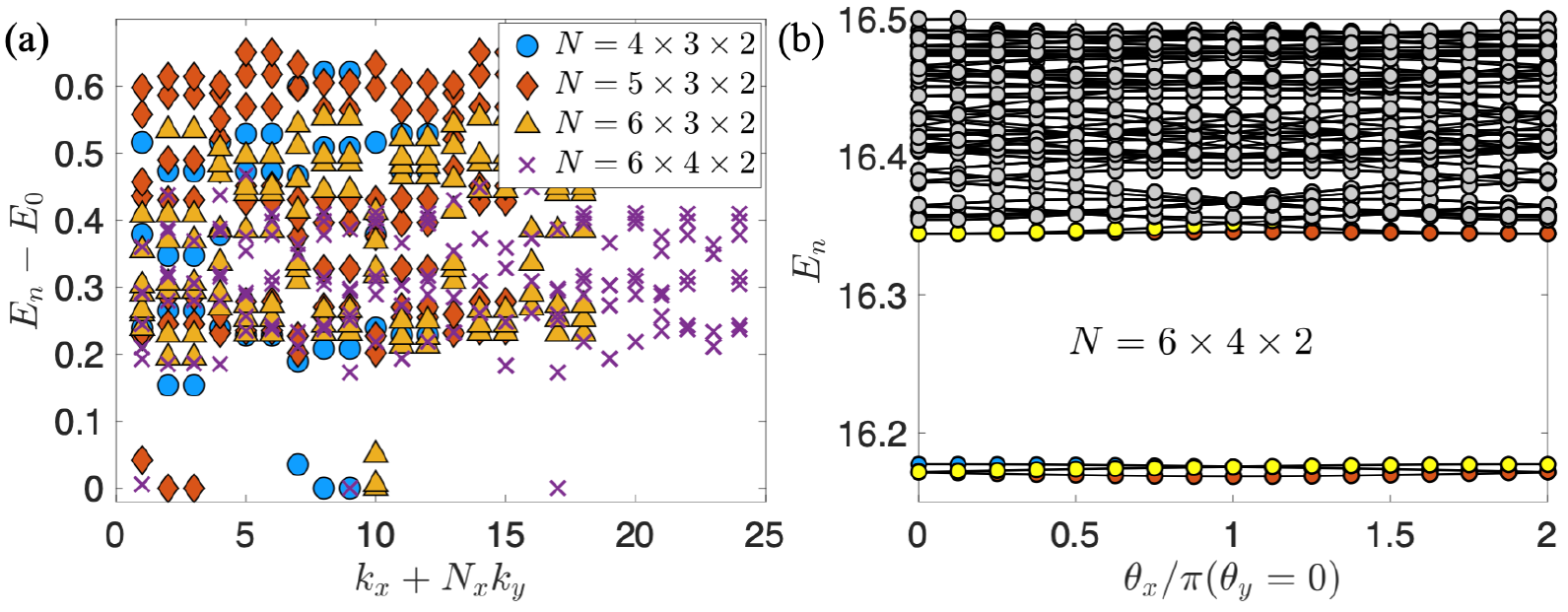}
		\caption{(a) The energy spectra at $V_1=4$. (b) Energy spectra flow with twisted boundary conditions.} 
		\label{figsm_ed_twist}
	\end{figure}
	
	\begin{figure}[htp!]
		\centering		
		\includegraphics[width=0.8\textwidth]{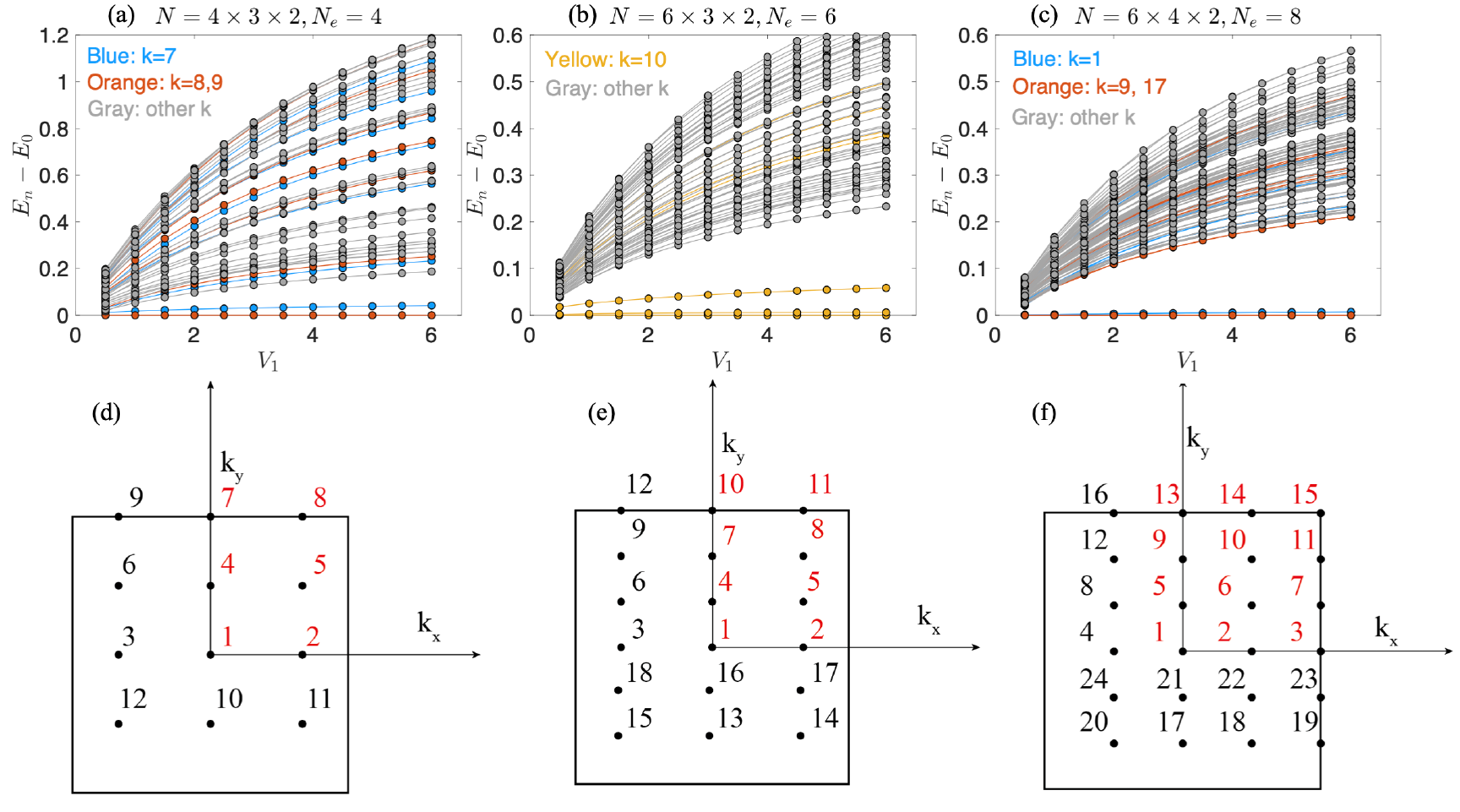}
		\caption{(a-c) The energy spectra with $V_1$ changing for $4\times3\times2$, $6\times3\times2$, $6\times4\times2$ torus respectively. (d-f) Corresponding momentum points in the Brillioun zone of unit cells. The energy spectra with red momentum sectors are calculated using Lanczos method, while the black ones can got by mirror or $C_4$ rotation symmetry.} 
		\label{figsm_ed_changeV}
	\end{figure}
	
	We have shown the $\nu=1/3$ energy spectra under different system sizes with $V_1=4$, and here we show the spectra of different system sizes changing with $V_1$ in Fig.\ref{figsm_ed_changeV}. It is in agreement with previous studies\cite{DNSheng2011_fci} that there is no phase transition in the large-$V_1$ region, and the unique feature of FQAH ground-state manifold is again exhibited that only in $3\times6\times2$ torus do the ground states appear in the same momentum sector. To characterized the topological ground state of FQAH, in ED calculation, we also employ twisted boundary condition (TBC) to calculate the Chern number of (quasi)degenerate groundstates using the following formula~\cite{Niu1985_hallconductance, Fukui2005_hallconductance, Varney2011_haldanemodel},
	\begin{equation}
		C=\frac{i}{2\pi}\int\int d\phi_1 d\phi_2\left[\frac{\partial}{\partial \phi_1}\langle \Omega(\phi_1,\phi_2)| \frac{\partial}{\partial \phi_2}| \Omega(\phi_1,\phi_2)\rangle -\frac{\partial}{\partial \phi_2}\langle \Omega(\phi_1,\phi_2)| \frac{\partial}{\partial \phi_1}| \Omega(\phi_1,\phi_2)\rangle\right], 
	\end{equation}
	where $\phi_1 \in [0, 2\pi)$ and $\phi_2 \in [0, 2\pi)$ are the twisted phases along $a_1$ and $a_2$ directions, respectively. In order to use the translational symmetry under TBC, These phases are evenly distributed across various bonds. The total Chern number of three (quasi)degenerate groundstates is equal to 1 no matter what the system size is. Therefore, we have numerically confirm the FQAH state at $\nu =1/3$ again.
	
	Except for the energy spectrum, we also calculated the density-density correlation function in the reciprocal space. For the $\nu=1/3$ FQAH state, there are some broad peaks at the edge of Brillouin zone as shown in Fig.\ref{figsm_ed_Sq}. The result of $N_s=6\times3\times2$ torus is similar to the XTRG and DMRG results with $3\times12\times2$ cylinder. The lack of $C_4$ symmetry in $S(\mathbf{q})$ is due to the geometry. When we change the torus to be $6\times4\times2$, as shown in Fig.\ref{figsm_ed_Sq}, it is $C_4$-symmetric, while the locations of broad peak shift along the boundary of Brillioun Zone. However, the common feature that the broad peaks are are the boundary with finite momentum will not change.
	
	\begin{figure}[htp!]
		\centering		
		\includegraphics[width=0.55\textwidth]{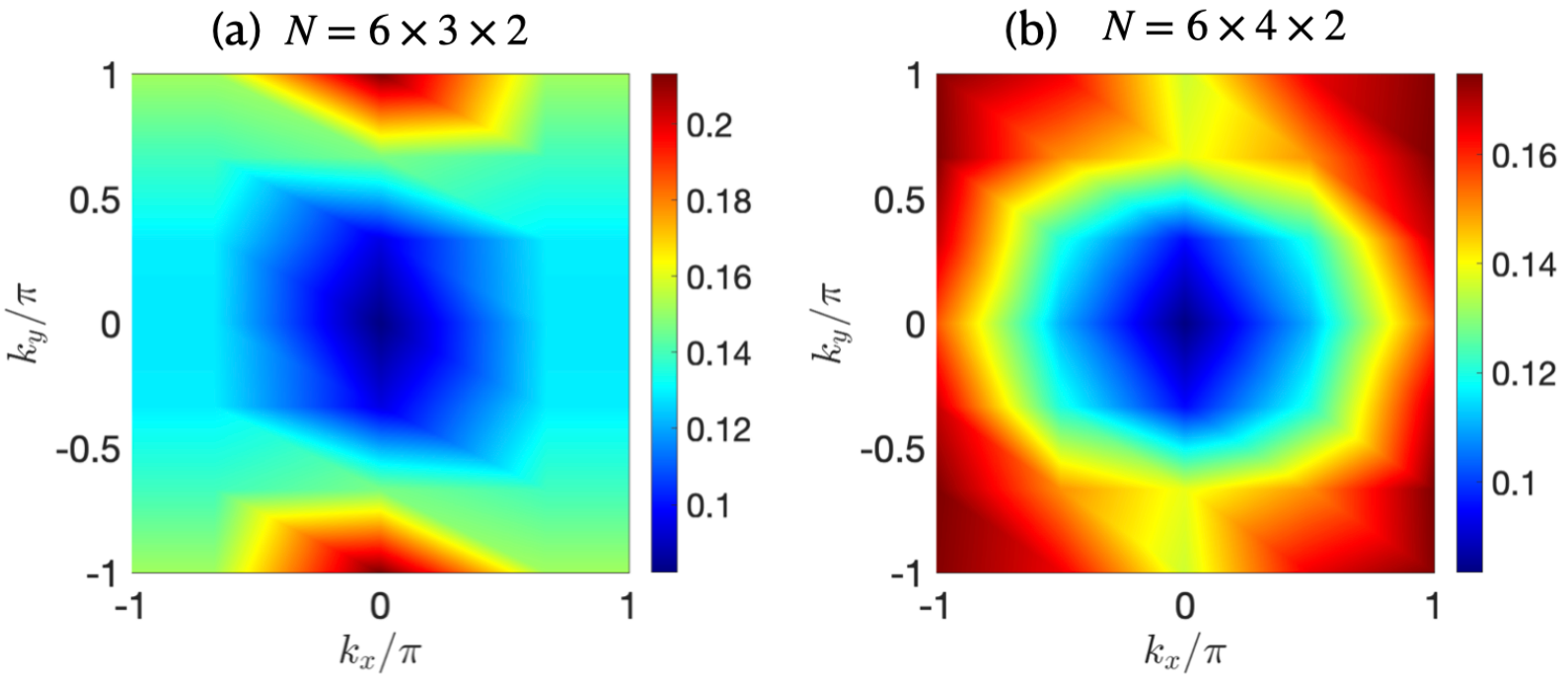}
		\caption{Structure factors of (a) $N=6\times3\times 2$ and $N=6\times4\times2$ respectively. The broad peaks are always at the boundary of the Brillioun zone. }
		\label{figsm_ed_Sq}
	\end{figure}
	
	\subsection{Section II:  DMRG results of $\nu=1/3$ structure factor}
	
	In this section, we also show the ground-state structure factors simulated using DMRG. The structure factor of $N=3\times12\times$ cylinder with $V_1=4$ is shown in Fig.\ref{figsm_dmrg_Sq}(a), which is the same geometry as the XTRG results in the main text. Here, the lack of $C_4$ symmetry is due to the odd $N_y$, which is similar to the $6\times3\times 2$ ED result. We also show the $Ny=3$ results with longer $N_x=24$ in Fig.\ref{figsm_dmrg_Sq}(b), and we plot the change of $S(\pi,0)$ versus $N_x$ in Fig.\ref{figsm_dmrg_Sq}(c) and find it almost constant.
	
	\begin{figure}[htp!]
		\centering		
		\includegraphics[width=\textwidth]{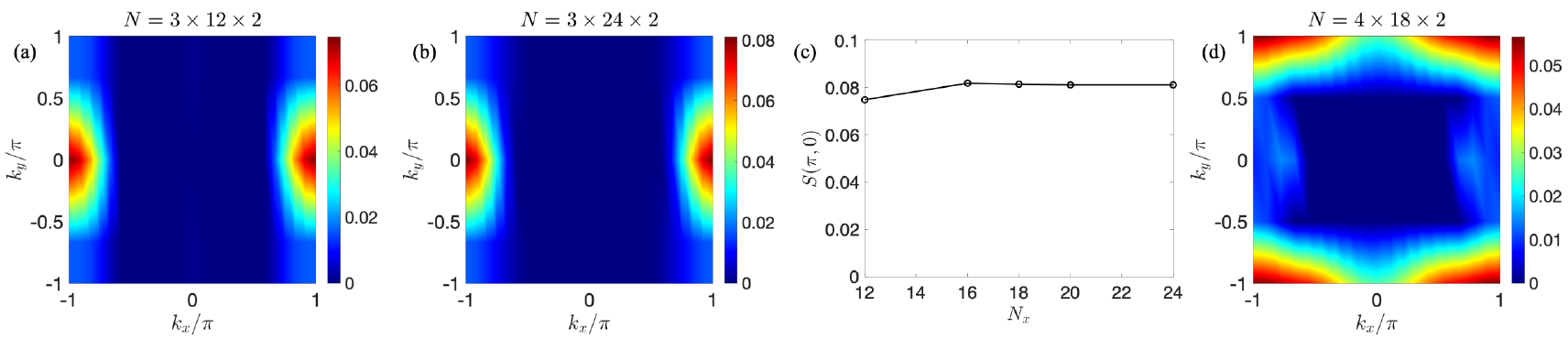}
		\caption{DMRG results of structure factors with $\nu=1/3$ filling and $V_1=4$. (a) $N=3\times12\times 2$ cylinder. (b) $N=3\times24\times2$ cylinder. (c) $S(\pi,0)$ with $N_y=3$ and changing $N_x$. (d) $N=4\times18\times2$ cylinder. }
		\label{figsm_dmrg_Sq}
	\end{figure}
	When we change the geometry to $N=4\times18\times2$ as shown in Fig.\ref{figsm_dmrg_Sq}(d), the structure factor is similar with the $N=6\times4\times2$ ED result, and the broad peaks shift towards the corner of the Brillion zone along the boundary. Again, the common feature is that the broad peaks are with finite momentum.
	
	\subsection{Section III: Exponential Tensor Renormalization Group Method}
	The main idea of exponential tensor renormalization group (XTRG) \cite{BBChen2018_XTRG} method is, to first construct
	the initial high-temperature density operator $\hat{\rho_0} \equiv \hat{\rho}(\tau ) = e^{-\tau H} $ with $\tau$ being an exponentially small inverse temperature, which can be obtained with ease via Trotter-Suzuki decomposition or series-expansion methods. Subsequently, we evolve the thermal state exponentially by squaring the density operator iteratively, i.e., $\hat{\rho}_n \cdot \hat{\rho}_n \equiv \hat{\rho}(2^n \tau ) \cdot \hat{\rho}(2^n\tau ) \rightarrow \hat{\rho}_{n+1}$. Following this exponential evolution scheme, one can significantly reduce the imaginary-time evolution as well as truncation steps, and thus can obtain highly accurate low-$T$ data in greatly improved efficiencies.
	
	In XTRG simulations, we measure an observable $\hat{O}$:
	\begin{equation}
		\langle \hat{O} \rangle\ (T)=\frac{\text{Tr}(\hat{\rho}(\frac{\beta}{2})\ \hat{O}\  \hat{\rho}(\frac{\beta}{2}))}{\text{Tr}(\hat{\rho}(\frac{\beta}{2})\hat{\rho}(\frac{\beta}{2}))}
	\end{equation}
	where we write the operator $\hat O$ as an MPO and inverse temperature $\beta=1/T$.
	
	When adapting XTRG to fermion systems, one should take care of the fermionic sign of exchanging two electrons. In this work, we are working on the many-body basis $| n_1\ n_2 \cdot\cdot\cdot n_N \rangle \equiv (c_N^\dagger)^{n_N}\cdot\cdot\cdot(c_2^\dagger)^{n_2}(c_1^\dagger)^{n_1}|\Omega \rangle$, where $n_i \in  \{ 0, 1 \}$ is the number of electrons at the site $i$ and $|\Omega\rangle$ is the vacuum state. Generically in this basis, the one-body operator $c_i^\dagger c_j$ (assuming $j < i$) requires an sign $\prod_{l=j+1}^{i} (-1)^{n_l}$, in addition to transform the state $|n_1 \cdot\cdot\cdot n_j \cdot\cdot\cdot n_i \cdot\cdot\cdot n_N \rangle $ to the state $|n_1 \cdot\cdot\cdot n_{j-1} \cdot\cdot\cdot n_{i+1} \cdot\cdot\cdot n_N \rangle $.

\end{widetext}

\end{document}